\documentstyle[graphicx,12pt]{article}
\begin{document}
\title{Application of Contractor Renormalization Group (CORE) to the Heisenberg zig-zag\\and the Hubbard chain}
\author{Krzysztof Cichy\footnote{Corresponding author; e-mail: krzystof.cichy@gmail.com}, Piotr Tomczak}
\date{}
\maketitle
\begin{center}
\vspace{-1cm}
\emph{Quantum Physics Division,\\ Faculty of Physics, Adam Mickiewicz University,\\ Umultowska 85,
61-614 Pozna\'n, Poland}
\end{center}


\begin{abstract}
The COntractor REnormalization group method was devised in 1994 by Morningstar and Weinstein.
It was primarily aimed at extracting the physics of lattice quantum field theories (like lattice Quantum Chromodynamics).
However, it is a general method of analyzing Hamiltonian lattice systems, e.g. Ising, Heisenberg or Hubbard models.
The aim of this work is to show the application of CORE to one-dimen\-sio\-nal (1D) quantum systems -- the Heisenberg zig-zag model and the Hubbard chain.
As a test of the method, the ground state energy of these systems will be calculated.
\end{abstract}

\section{Introduction}
The COntractor REnormalization group (CORE) method was invented by Morningstar and Weinstein \cite{morn1994}, \cite{morn1996} as an alternative to other real-space renormalization group methods and Monte Carlo techniques for the Feynman path integral evaluation.
The computation with the use of this method begins by restricting the full Hilbert space of a single block to an appropriately chosen subspace.
Then, clusters of two and more blocks are considered.
The overlap of the lowest lying eigenstates with the tensor products of the retained single-block eigenstates is calculated, which allows to construct a renormalized Hamiltonian.
Then, one can iterate the renormalization group procedure (which is straightforward if the renormalized Hamiltonian has the same form as the original one) or use some other methods to examine the properties of the effective (renormalized) Hamiltonian.

So far, the CORE framework has been used mainly to analyze the Heisenberg model \cite{piekarewicz1997}, \cite{piekarewicz1998}, \cite{wein2001}, \cite{berg}, \cite{budnik}, \cite{capponi2004}, \cite{li2004}.
Other applications included the 2D Hubbard model \cite{altman2002} and the 2D $t$-$J$ model \cite{capponi2002}.

In this paper we will apply CORE to the 1D Heisenberg model with nearest-neighbor and next-nearest-neighbor interactions (the Heisenberg zig-zag, also called the Majumdar-Ghosh model) and to the Hubbard chain in the cases of both on-site attraction and repulsion.
The paper is organized as follows.
Section 2 briefly summarizes the CORE method.
Section 3 deals with the Heisenberg zig-zag.
In Section 4 we move on to the Hubbard chain.
Section 5 concludes.

\section{Basics of CORE}
We implement the following algorithm for computations (in one dimension):
\begin{enumerate}
\item Divide the lattice into disjoint and identical blocks of $L$ sites.
\item Diagonalize the single-block Hamiltonian $H_B$ and keep the $M$ lowest lying eigenstates.
\item Construct the projection operator $P$ on the subspace of the retained eigenstates.
Range-1 renormalized Hamiltonian for block $j$ is defined by $H^{ren}_1(j)=PH_B(j) P\equiv h_1(j)$.
$h_1(j)$ denotes the so-called range-1 term in the cluster expansion of the renormalized Hamiltonian.
\item Consider a cluster of $r$ connected blocks. 
Construct the reduced Hilbert space which is spanned by $r$ tensor products of the retained $M$ states.
Diagonalize the cluster Hamiltonian and keep its $M^r$ lowest lying eigenstates\footnote{In case of degeneracies in the spectrum of the cluster Hamiltonian one sometimes has to keep more than $M^r$ eigenstates at this stage and at a later stage perform a singular value decomposition. The full algorithm one has to follow then is described in detail in \cite{morn1996}.} $|\psi_i\rangle$, where $i=1,\ldots,M$.
\item Project the $M^r$ states on the reduced Hilbert space, obtaining a set of $M^r$ wave functions, which are to be Gram-Schmidt orthonormalized.
At the end of this procedure one has a set of $M^r$ states $|\tilde{\psi}_i\rangle$.
\item Define the range-$r$ renormalized Hamiltonian:
\begin{equation}
{H}^{ren}_r \equiv \sum_n^{M^r} \varepsilon_n |\tilde{\psi}_n\rangle\langle\tilde{\psi}_n|,
\end{equation}
where $\varepsilon_n$ denotes the eigenenergy of the $n$-th lowest lying eigenstate of the cluster Hamiltonian.
\item Range-$r$ term in the cluster expansion of the renormalized Hamiltonian is given by:
\begin{eqnarray}
h_r(j,\ldots,j+r-1)&=&H_r^{ren}(j,\ldots,j+r-1)+\\
&-&\sum_{n=1}^{r-1} \sum_{m=0}^{r-n} h_n(j+m,\ldots,j+n+m-1),\nonumber
\end{eqnarray} 
i.e. one subtracts from the range-$r$ Hamiltonian all of the terms already included in the range-$n$ ($n<r$) computations.
\item Repeat steps 4-7 for more connected blocks.
Neglect terms including more than some specified value $r_{max}$ connected blocks, depending on the range of correlations in the original Hamiltonian.
Usually, taking terms of range 2, 3 or 4 is enough to extract the quantities of interest.
\item The infinite-lattice renormalized Hamiltonian is the sum of terms of range from 1 to $r_{max}$:
\begin{equation}
H^{ren}=\sum_{j=-\infty}^{\infty} \sum_{r=1}^{r_{max}} h_{r}(j,\ldots ,j+r-1).
\end{equation}
\end{enumerate}

\section{The Heisenberg zig-zag}
In this section we will use CORE to calculate the ground state energy of an infinite Heisenberg chain with nearest-neighbour (n.n.) and next-nearest-neighbour (n.n.n.) interactions.
The model Hamiltonian is:
\begin{equation}
H=\sum_i \mathbf S\mathnormal_i\cdot\mathbf S\mathnormal_{i+1} + J\sum_i \mathbf{S}\mathnormal_i\cdot\mathbf{S}\mathnormal_{i+2},
\end{equation} 
where $\mathbf{S}\mathnormal_i$ is site-$i$ spin-1/2 operator and $J$ denotes the ratio of exchange integrals for n.n.n. and n.n. interactions.

We start with calculations for $M=1$ retained states of the single-block Hamiltonian with a single block consisting of $L=2$ (scheme A) and $L=3$ (scheme B) spins.
Then, range-$r$ renormalized Hamiltonian equals the ground state energy of $r$ connected blocks.
The range-$r$ terms in the cluster expansion equal: $h_1=H^{ren}_1$, $h_2=H^{ren}_2-2h_1$, $h_3=H^{ren}_3-2h_2-3h_1$, $h_4=H^{ren}_4-2h_3-3h_2-4h_1$ etc.
The range-$r$ ground state energy estimate is given by the sum of the first $r$ terms $h_n$.

\begin{figure}[t]
\begin{center}
\includegraphics[width=0.34\textwidth,angle=270]{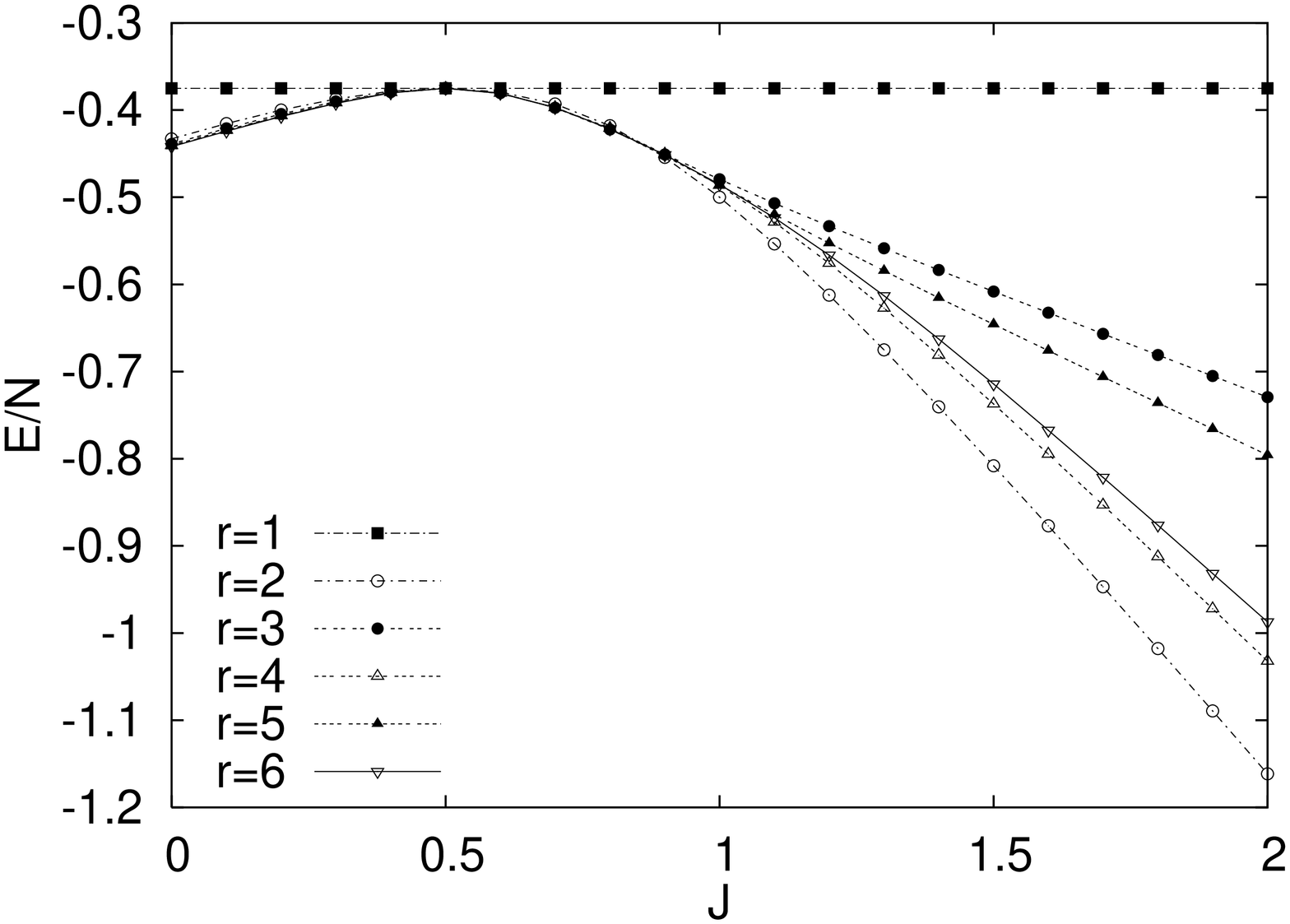}
\includegraphics[width=0.34\textwidth,angle=270]{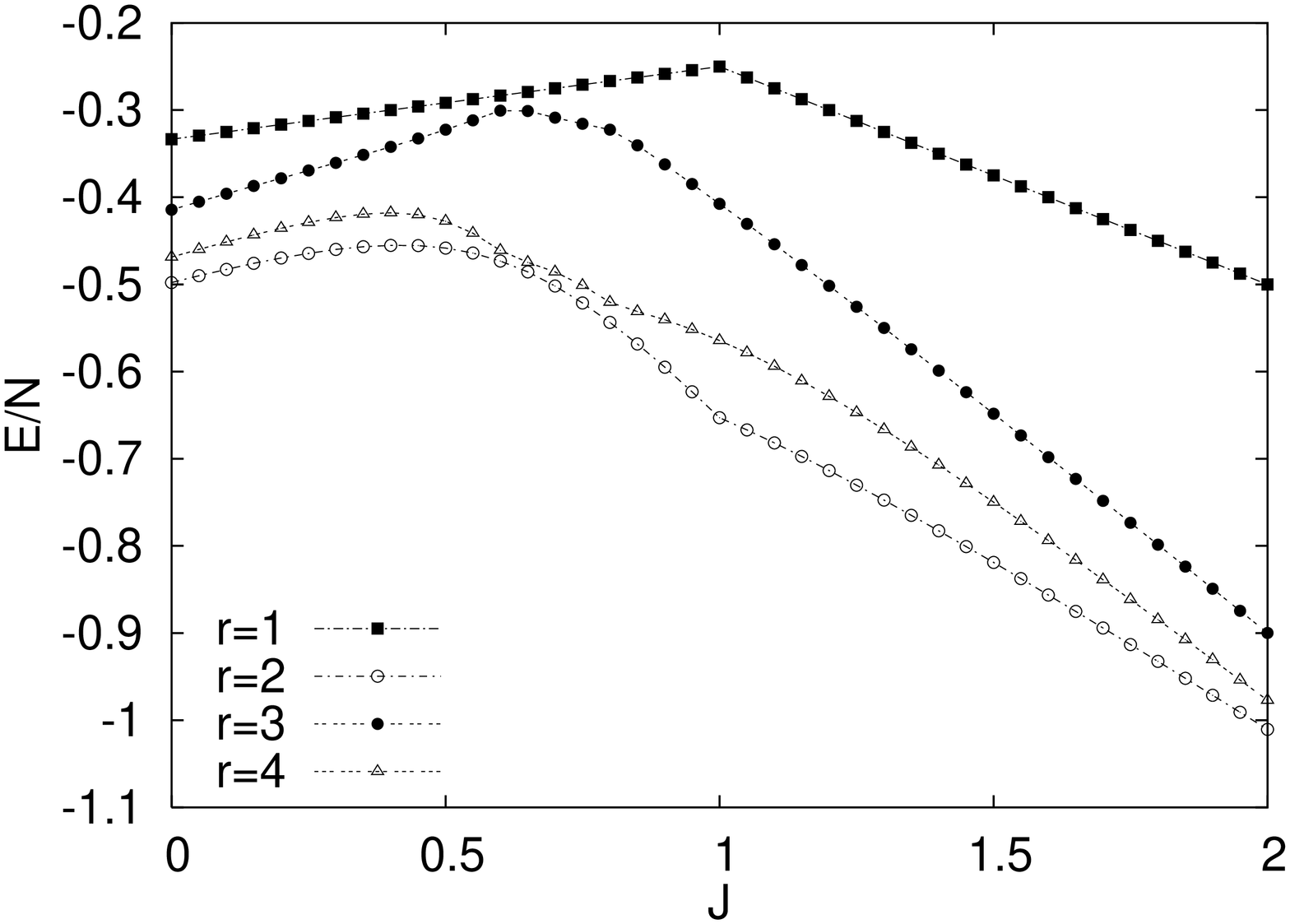}
\caption{(left) The range-1 to range-6 estimates of the ground state energy per spin for the Heisenberg zig-zag (scheme A: $L=2$ spins per block, $M=1$ retained state).
(right) The range-1 to range-4 estimates of the ground state energy per spin for the Heisenberg zig-zag (scheme B: $L=3$ spins per block, $M=1$ retained state).}
\end{center}
\end{figure}

Figure 1 (left) shows the computations of range-1 to range-6 estimates for the ground state energy for scheme A.
The range-6 computation requires the exact diagonalization of $4096\times 4096$ matrices for twelve-spin blocks which is quite intensive in terms of computer time and memory, but still feasible even on a PC.
Each further block takes a factor of 64 of computer time and a factor of 16 of computer memory more, rendering calculations for bigger blocks very hard.
The range-1 approximation does not depend on $J$, since within a two-spin block the n.n.n. interaction is not possible.
The case of $J=0$ corresponds to the Heisenberg chain, for which an exact result of Bethe \cite{bethe} and Hulth\'en \cite{hulthen} is known ($1/4-\ln2\approx-0.443147$).
The CORE range-6 result is -0.442028, which is just 0.25\% above the exact result.
The precision of this result is impossible to obtain within other approximation schemes, like the ''naive'' renormalization group or spin-wave approximation.
The case of $J=0.5$ is the Majumdar-Ghosh limit, in which the ground state is a set of non-interacting singlets.
CORE reproduces this result in range-1 computation, i.e. higher ranges are identically zero.
For $J\in[0,1]$ range-5 and range-6 terms are very small.
This shows that the range-6 result is very near to the exact answer, with an estimation error of the same range as for the $J=0$ case.
For $J>1$, the CORE ground state energy estimates begin to diverge with the exact result being somewhere between the range-5 and range-6 curve.
Therefore, CORE, at least for the $L=2$ scheme, seems to be unreliable for $J>1$.
However, for $J<1$, the results are very accurate.

Figure 1 (right) shows the computations of range-1 to range-4 estimates for the ground state energy for scheme B.
In this case, the three-spin and nine-spin block ground states are doubly degenerate, while the six-spin and twelve-spin block ground states are non-degenerate.
This difference in the block ground state structure accounts for the fact that this approximation scheme fails for all values of $J$, despite the fact that the range-4 computation is as time- and memory-consuming as the range-6 calculation within scheme A.
This example suggests that an appropriate blocking procedure is essential in CORE.

The block ground state structure gives a hint of such procedure for the case of $L=3$ -- one should keep (at least) two, instead of one, single-block states to construct the reduced Hilbert space.
Such scheme ($M=2$ retained states of a single-block ($L=3$) Hamiltonian) will be now used.
In this case, the range-2 renormalized Hamiltonian for block $j$ takes the form:
\begin{equation}
H^{ren}_2(j)=C_2\mathbf{1}\mathnormal_j+\sum_{i,i+1\in j}\alpha_2 \left(\mathbf S\mathnormal_i\cdot \mathbf S\mathnormal_{i+1}\right),
\end{equation} 
where the parameters $C_2$ and $\alpha_2$ are to be found.
The full (infinite-lattice) range-2 renormalized Hamiltonian is the sum of all range-1 and range-2 terms:
\begin{equation}
H^{ren}=\sum_{i=-\infty}^{\infty}\Big( (C_2+1) \mathbf{1}\mathnormal_i+\alpha_2 \left(\mathbf S\mathnormal_i\cdot \mathbf S\mathnormal_{i+1}\right)\Big).
\end{equation} 
To obtain the estimate for the ground state energy, one has to iterate this Hamiltonian infinitely many times.
However, the iterations preserve this form of the Hamiltonian.
After $n$ iterations, the renormalized Hamiltonian is:
{\setlength \arraycolsep{1pt}
\begin{eqnarray}
H^{ren-n}&=&\sum_{i=-\infty}^{\infty}\Bigg((\alpha'_2)^{n-1}\alpha_2 \mathbf \left(S\mathnormal_i\cdot \mathbf S\mathnormal_{i+1}\right)\mathnormal+\Big(3^{n-1}(C_2+1)+3^{n-2}\alpha_2(C'_2+1)\nonumber\\
&&+\ldots+3\alpha_2(\alpha'_2)^{n-3}(C'_2+1)+\alpha_2(\alpha'_2)^{n-2}(C'_2+1)\Big)\mathbf{1}\mathnormal_i\Bigg),
\end{eqnarray}}
where $C'_2=-2.124893$ and $\alpha'_2=0.491582$ denote the parameters of renormalization for the $J=0$ Heisenberg chain.
Thus, the ground state energy per spin is equal to the $n\rightarrow\infty$ limit of the expression in brackets, divided by the volume of the lattice.
Taking into account that we have obtained an expression for an infinite geometric series, we finally get:
\begin{equation}
E/N=\frac{C_2+1}{3}+\frac{\alpha_2(C'_2+1)}{(9-3\alpha'_2)}.
\end{equation}
For the $J=0$ case, this expression reduces to the Morningstar-Weinstein result \cite{morn1996} for the range-2 estimate of the ground state energy of the Heisenberg chain ($(C'_2+1)/(3-\alpha'_2)=-0.448446$), just 1.2\% below the exact result of Bethe and Hulth\'en.

The range-3 renormalized Hamiltonian for block $j$ takes the form:
\begin{equation}
H^{ren}_3(j)=C_3\mathbf{1}\mathnormal_j+\sum_{i,i+1\in j}\alpha_3 \left(\mathbf S\mathnormal_i\cdot \mathbf S\mathnormal_{i+1}\right)+\gamma_3 \left(\mathbf S\mathnormal_{i\in j}\cdot \mathbf S\mathnormal_{i+2\in j}\right).
\end{equation} 
The full (infinite-lattice) range-3 renormalized Hamiltonian is the sum of all \mbox{range-1}, range-2 and range-3 terms:
\begin{equation}
H^{ren}=\sum_{i=-\infty}^{\infty}\Big((C_3-C_2)\mathbf{1}\mathnormal_i +(2 \alpha_3-\alpha_2) \left(\mathbf S\mathnormal_i\cdot \mathbf S\mathnormal_{i+1}\right)+\gamma_3 \left(\mathbf S\mathnormal_i\cdot \mathbf S\mathnormal_{i+2}\right)\Big).
\end{equation} 
To obtain the estimate for the ground state energy per spin, one again has to iterate $H^{ren}$ infinitely many times (the form of the Hamiltonian is preserved in successive iterations).
Again, the factor multiplying the identity operator after infitely many iterations gives the estimate of the ground state energy.
For the $J=0$ case, the estimate agrees with the Weinstein result \mbox{-0.447635} \cite{wein2001}.

\begin{figure}
\begin{center}
\includegraphics[width=0.34\textwidth,angle=270]{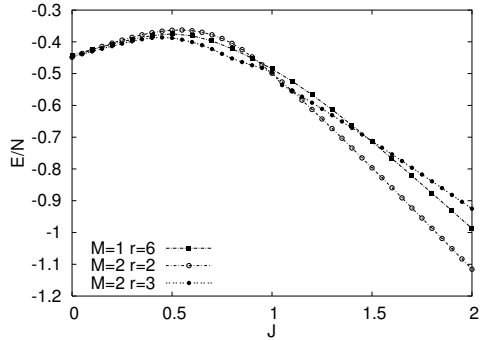}
\caption{The range-2 and range-3 estimates of the ground state energy per spin for the Heisenberg zig-zag ($L=3$ spins per block, $M=2$ retained states). The range-6 estimate ($L=2$, $M=1$) given for comparison.}
\end{center}
\end{figure}

Figure 2 shows the computations of range-2 and range-3 estimates for the ground state energy per spin.
The dashed line corresponds to the value for range-2 and the dotted line for range-3.
The solid line is the case of range-6, $L=2$, $M=1$ computation (scheme A), which is given for comparison (for $J<1$ it corresponds nearly to the exact result).
For the values of $J<0.5$, all three lines almost coincide.
The range-3 correction to the energy is small in this region, which is again the result of a different srtucture of the ground state for six-spin and nine-spin clusters.
Weinstein argues for the case of $J=0$ that the range-4 term should be much bigger than the range-3 term and shows this explicitly \cite{wein2001}.
One can expect that his argument holds for the nonzero-$J$ case, so that performing the range-4 computation should give a more exact result (closer to the range-6 approximation in scheme A).
For $J\in[0.5,1)$, the range-2 and range-3 results differ visibly from each other and from the scheme A prediction.
One can assume that again the range-4 computation should fix this problem.
For $J$ slightly below 1, the range-2 and range-3 lines almost meet (the jump of the ground state energy estimate for $J=1$ is a result of the fact that the ground state energy of a nine-spin block is four-times degenerate in this case) and for $J>1$ one can observe an ever-increasing difference between the lines, similarly to the scheme A case for $J>1$.
In this regime, CORE seems to fail in all of the schemes under consideration.

\section{The Hubbard chain}
In this section, we will show simple CORE calculations of the ground state energy per site of the Hubbard chain at half-filling with both attractive and repulsive on-site interactions.
We will retain only the ground state of block and cluster Hamiltonians ($M=1$), so that the renormalized Hamiltonians will be real numbers.
A single block will consist of two lattice sites and we will consider clusters of at most 6 sites (the numerical form for the six-site Hamiltonian is already a $924\times924$ matrix).
The relevant formulas for $h_n$ ($n=1,2,3$) are the same as in the Heisenberg zig-zag case.

The model Hamiltonian is:
\begin{equation}
H=-t\sum_{<i,j>,\sigma}\left(c_{i\sigma}^\dagger c_{j\sigma}+H.c.\right)+U\sum_i n_{i\uparrow} n_{i \downarrow},
\end{equation} 
where $t$ is the hopping integral, $c_{i\sigma}$ and $c^\dagger_{i\sigma}$ are annihilation and creation operators for an electron of spin $\sigma$ at site $i$, $U$ is the on-site interaction strength and $n_{i\sigma}$ are the number operators of electrons of spin $\sigma$ at site $i$.
The sum extends over nearest neighbours.

The exact solution for the ground state energy per site in the 1D case with on-site repulsion was given by Lieb and Wu \cite{lieb-wu}:
\begin{equation}
\frac{E^{rep}(U)}{N}=-4\int_0^\infty dx \frac{J_0(x)J_1(x)}{x(1+e^{xU \slash 2})},
\end{equation} 
where $J_0$ and $J_1$ are Bessel functions of zero and first order.

In the case of on-site attraction, the relevant expression is:
\begin{equation}
\frac{E^{attr}(U)}{N}=\frac{U}{2}+\frac{E^{rep}(|U|)}{N}.
\end{equation}

\begin{figure}[t]
\begin{center}
\includegraphics[width=0.34\textwidth,angle=270]{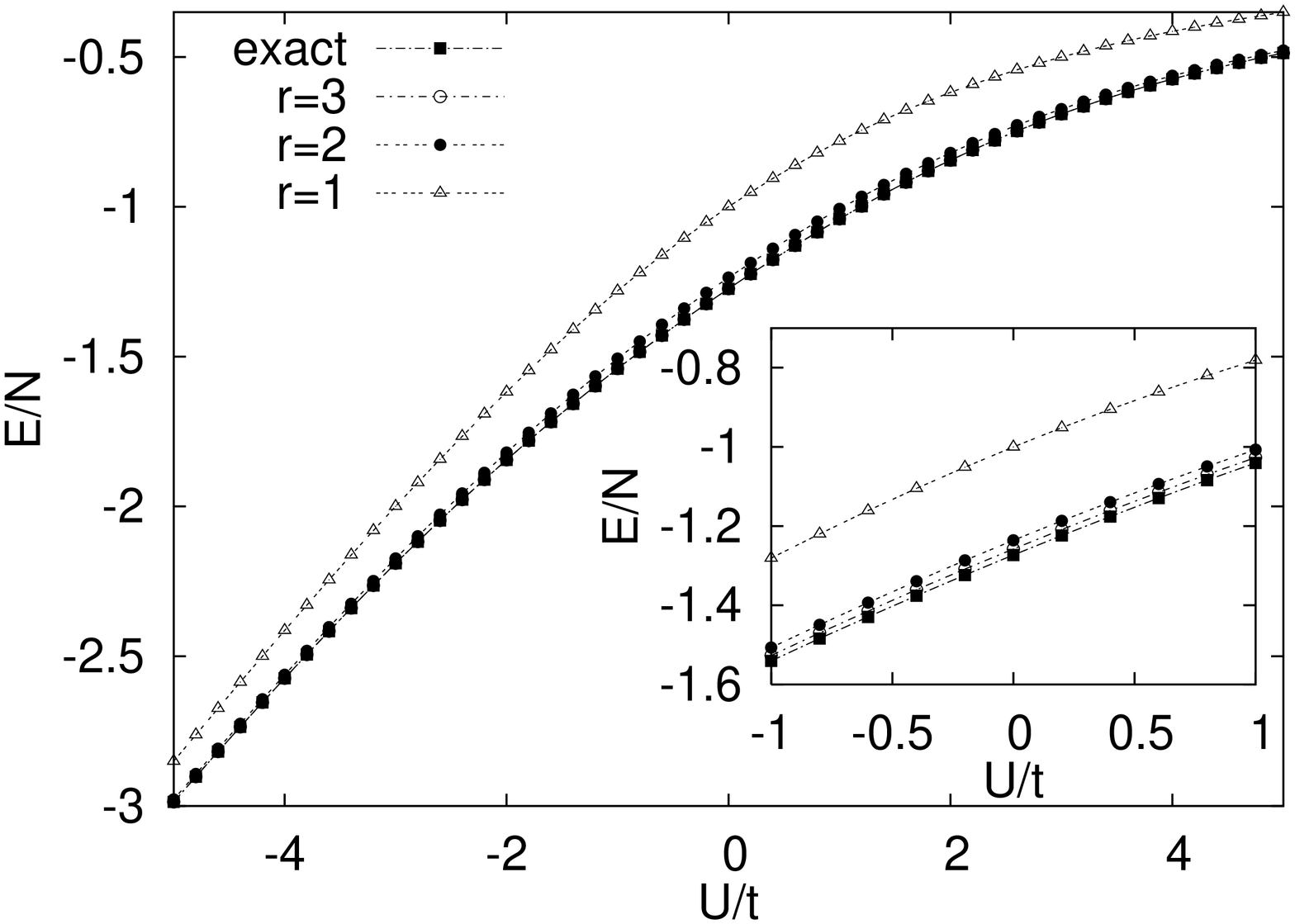}
\includegraphics[width=0.34\textwidth,angle=270]{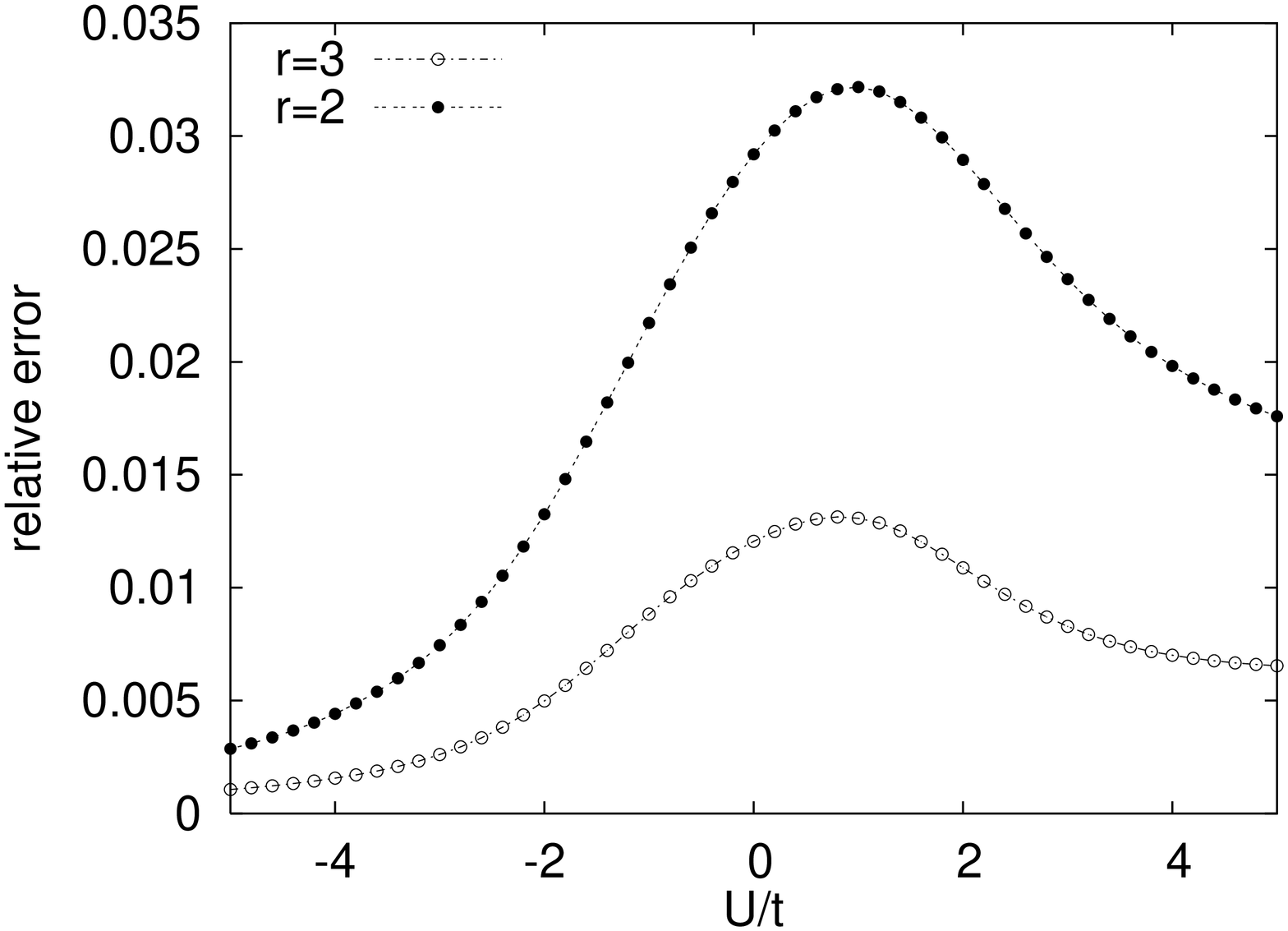}
\caption{(left) The range-1 to range-3 estimates of the ground state energy and the exact result of Lieb and Wu for the Hubbard chain per site at half-filling ($L=2$ sites per block, $M=1$ retained state).
(right) The errors of estimation of the ground state energy per site for the Hubbard chain, relatively to the exact result.}
\end{center}
\end{figure}

Figure 3 (left) shows the computations of range-1 to range-3 estimates for the ground state energy per site vs. interaction strength $U/t$.
The relative errors of these calculations are given on the right side of Figure 3.
One can see that the biggest error corresponds to the case of $U/t\approx1$ and is of the order of 1\% (range-3) and 3\% (range-2).
For a bigger repulsion strength and especially in the attractive case the estimation error is much smaller (only around 0.1\% for $U/t\approx-5$).
This results from the fact that the relatively small clusters that we consider seem to be enough to capture the case when the probability that electrons are close to one another is big.
At the same time, when the tendency to localization and delocalization is comparable ($U\approx t$), blocks of this size are too small to accurately describe such case.

However, the relatively small divergence from the exact Lieb-Wu result seems to show that the mechanisms relevant for the Hubbard chain at half-filling take place at distance scales of not much more than six neighbouring sites.

\section{Conclusions}
In this paper we have shown that the CORE method works quite well for 1D quantum systems -- the Heisenberg zig-zag and the Hubbard chain at half-filling.
The method was tested in calculation of the ground state energy per lattice site and the results seem to be encouraging.
A further test of the method would be to calculate other operators of interest in the CORE framework and compare it with available exact or Quantum Monte Carlo results.

\end{document}